\documentclass[a4paper,twoside,12pt]{article} 
\usepackage{amsmath}
\usepackage{amssymb}
\usepackage[OT4]{fontenc}
\usepackage[english,polish]{babel}
\usepackage[cp1250]{inputenc}
\usepackage[MeX]{polski}
\usepackage{graphicx}
\usepackage{here}
\usepackage{anysize}   
\usepackage{indentfirst} 
\usepackage{chngcntr} 
\counterwithin{figure}{section}
\counterwithin{equation}{section}
\counterwithin{table}{section}
\marginsize{3,0cm}{3,0cm}{2,5cm}{2,5cm} 
\usepackage{fancyhdr}

\begin{document}

\selectlanguage{english}

\begin{titlepage} 

\begin{center}
\large  {
INSTITUTE OF PHYSICS\\
FACULTY OF PHYSICS, ASTRONOMY\\ 
AND APPLIED COMPUTER SCIENCE\\
JAGIELLONIAN UNIVERSITY }
\end{center}	

\vspace{1.2cm}

\begin{center}
	\Large \textbf{Feasibility study of measuring CP symmetry violation via $\eta \to 4\pi$ decay using WASA-at-COSY detector}
\end{center}

\vspace{2.0cm}

\begin{center}
\Large \textbf{Tomasz Bednarski}
\end{center}

\vspace{1.0cm}

\begin{center}
\large{Diploma Thesis}
\end{center}

\vspace{0.3cm}

\begin{center}

supervised by:\\

\vspace{0.3cm}

\large{ Prof. dr hab. Pawe\l~Moskal}

\vspace{1.7cm}

\begin{figure}[h]
\begin{center}
\includegraphics[scale=0.4]{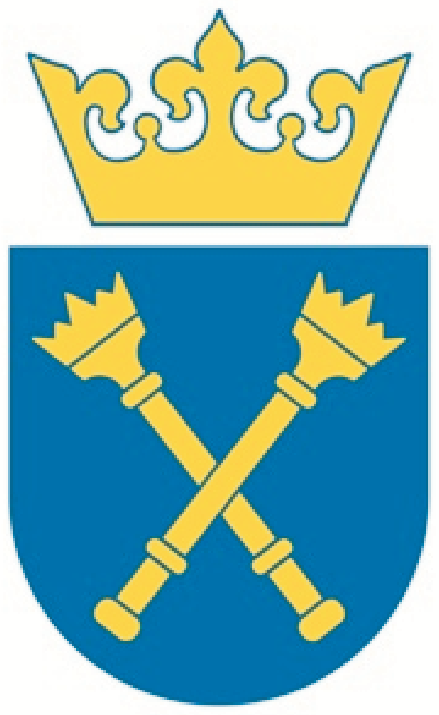}
\end{center}
\end{figure}

\vspace{0.2cm}

Cracow, 2011

\end{center}

\end{titlepage}

%%%%%%%%%%%%%%%%%%%%%%%%%%%%%%%%%%%%%%%%%%%%%%%%%%%%%%%%%%%%%%%%%%%%%%%%%%%%%%%

\newpage
\thispagestyle{empty}
\begin{center}\end{center}
\newpage
\thispagestyle{empty}
\begin{center} 
\vspace{2.cm}
\Large{\bf Abstract} 
\end{center}

\vspace{1cm}

It is known that the Standard Model does not describe all phenomena related to subatomic particles. This thesis presents feasibility studies of the measurement of $\eta$ meson decay using \mbox{WASA-at-COSY} detector which tests predictions of the Standard Model. The aim of this thesis is to estimate the time of measurement for which the current branching ratio upper limit of $\eta \to 4\pi^{0}$ decay can be improved.

In order to estimate the time of measurement Monte Carlo simulations were performed. Kinematics of the $pp \to pp\eta \to pp4\pi^{0} \to pp8\gamma$ reactions was simulated using PLUTO program. Next, by means of the GEANT3 program the response of the \mbox{WASA-at-COSY} detector was simulated for each particle. Received signals were analysed in RootSorter software package based on ROOT.	

Studies of the reaction with many gamma quanta in the exit channel required the investigation of \mbox{WASA-at-COSY} calorimeter functionality. Tests of a cluster building algorithm with emphasis on merging and splitting of detected signals were done.

The obtained result shows that only about 3\% of $\eta \to 4\pi^{0}$ decays can be properly reconstructed. The identification of the $\eta \to 4\pi^{0} \to 8\gamma$ decay was based on the invariant mass method for both $\pi^{0}$ and $\eta$ mesons. As a result, the resolution of the invariant mass determination of the $\eta$ meson was established to $\sigma = 31\ MeV/c^{2}$. The acceptance of the \mbox{WASA-at-COSY} detector for $pp \to pp\eta \to pp4\pi^{0} \to pp8\gamma$ reaction was determined to $A_{\eta}=N_{\eta\ detected}/ N_{\eta\ produced} = 1.5\%$. The time of measurement for obtaining a statistical precision equal to the current branching ratio upper limit for $T_{beam}=1.3\ GeV$ is about $13$ hours. It has also been established that the lower the excess energy in the $\eta$ production is the shorter the time of the measurement needed.

\newpage
\thispagestyle{empty}
\begin{center}\end{center}
\newpage

\tableofcontents  

\newpage
\thispagestyle{empty}
\begin{center}\end{center}
\newpage
%%%%%%%%%%%%%%%%%%%%%%%%%%%%%%%%%%%%%%%%%%%%

\pagestyle{fancy}
\fancyhf{} 
\fancyhead[LE,RO]{\textbf{\thepage}}
\fancyhead[RE]{\small\textbf{{Introduction}}} 
\newpage
\thispagestyle{plain}

\section{Introduction} \label{sec:introduction} 

\vspace{0.5cm}

Since last century the concept of symmetry has become very important in physics. The 1918 year, in particular and famous Noether’s theorem, which relates symmetries to conservations laws is arguably considered as a crucial moment in the history of the concept of symmetry. The first one who regarded symmetry principle as the primary feature of nature was Albert Einstein. He recognized the symmetry implicitly in Maxwell's equations and elevated it to a symmetry of space-time itself~\cite{Gross}.	

The symmetries have found applications in the Standard Model (SM) which is a theory describing the electromagnetic, weak, and strong interactions, which govern the dynamics of the known subatomic particles. However, although almost all experimental data can be explained by the SM, it does not explain basic phenomena such as the existence of three families of fundamental fermions, the origin of CP violation, etc. It is expected that the SM breaks down at some point. A place where New Physics can be looked for is the limit of validity of the basic symmetries of charge conjugation (C), parity (P) and time reversal (T), as well as combination of these symmetries CP and CPT~\cite{Nefkens}.

In the Standard Model CP violation is described by the phase in the Cabibbo-Kobayashi-Maskawa quark-mixing matrix. Six quark flavours are grouped into three families. CP violation is related to family-changing interactions, while in family-conserving cases CP violation is not included in the SM. It is deemed that detailed studies of CP violation may lead us to New Physics that goes beyond the Standard Model. Decays which test flavour-conserving P as well as CP violation are $\eta$ and $\eta'$ decays into $2\pi$~\cite{Nefkens}.

$\eta$ meson is a part of the light pseudoscalar nonet together with mesons $\pi,\ K$ and $\eta'$. Real $\eta$ and $\eta'$ mesons are a combination of the octet and singlet states ($\eta_{8}$ and $\eta_{1}$) of the SU(3) symmetry.
The $\eta$ meson is eigenstate of $\hat{C},\ \hat{P}$ and $\widehat{CP}$ operators, $C(\eta) = +1,\ P(\eta)=-1$. Therefore, the decays of the $\eta$ meson constitutes a tool to study these symmetries. The decay of $\eta \to 2\pi^{0}$ is the simplest example of a decay violating P and CP symmetry. The branching ratio for this decay is predicted to be less than $6\cdot 10^{-16}$ \cite{Kupsc:2011ah} and a present experimental upper limit amounts to $3.5 \cdot 10^{-4}$~\cite{PDG}. 

CP symmetry test $\eta \to 2\pi^{0}$ is hard to improve due to the background connected with direct $2\pi^{0}$ production in every $\eta$ production reaction. A possible new test is the decay into four pions~\cite{Nefkens}:

\begin{equation*}
	\eta \to 4\pi^{0}
	\label{p_so4p_eq1}
\end{equation*}

The present upper limit for branching ratio $BR(\eta \to 4\pi^{0}) < 6.9\cdot 10^{-7}$ was determined by Crystal Ball Collaboration \cite{CrystalBall}. 

CP symmetry in $\eta \to 4\pi^{0}$ can be conserved if wave function of pions in final state involves higher partial waves. However, in this case, estimated branching ratio of the CP conserving decay is less than $10^{-10}$ \cite{Kupsc:2011ah}.
A measurement of a non-zero BR, for this decay, above estimated limit would be a signal for a CP violating process. 

The main goal of this thesis is to estimate time of measurement, with \mbox{WASA-at-COSY} detector, for the reaction $pp \to pp\eta \to pp4\pi^{0} \to pp8\gamma$ needed to obtain precision better than a present upper limit of $BR(\eta\to 4\pi^{0})$.

Next section of this thesis comprises short theory introduction of symmetries in general with emphasis on parity and charge conjugation symmetry. The CP symmetry violation in $\eta$ decays ($\eta \to 2\pi^{0}$ and $\eta \to 4\pi^{0}$) are described in section~\ref{sec:CPV}. In section~\ref{reactions} the cross section for the $\eta$ meson production as well as the multipion production in proton-proton collision are presented. Further on electromagnetic calorimeter and other \mbox{WASA-at-COSY} detectors are described in section~\ref{sec:WASA}. The section~\ref{sec:kinematics} presents geometrical acceptance of the calorimeter and its efficiency for gamma quanta reconstruction. Level of statistical uncertainty of the branching ratio, achievable with \mbox{WASA-at-COSY} detector is determined as a function of time in section~\ref{sec:time}. Finally, the time of measurement needed to achieve a required  accuracy in the studies of $\eta \to 4\pi^{0}$ decay is estimated as a function of excess energy. 
	
The work is supplemented with Appendix where results of kinematical fit used for $pp \to pp\eta \to pp4\pi^{0} \to pp8\gamma$ reaction are presented.

%%%%%%%%%%%%%%%%%%%%%%%%%%%%%
\newpage
\thispagestyle{plain}

\pagestyle{fancy}
\fancyhf{} 
\fancyhead[LE,RO]{\textbf{\thepage}}
\fancyhead[RE]{\small\textbf{{Symmetries}}} 

\newpage

\section{Symmetries} \label{sec:symmetries}

\vspace{0.5cm}

Symmetries can be divided into continuous and discrete:

\begin{itemize}
	\item Continuous symmetries lead to conservation laws:
	\begin{enumerate}
		\item invariance under spatial transformation leads to momentum conservation
		\item invariance under time transformation leads to energy conservation
		\item invariance under spatial rotation leads to angular momentum conservation
		\item invariance under rotation in isospin space leads to isospin conservation
		\item invariance under a change in phase of a wave function of a charged particle leads to electric charge conservation 
	\end{enumerate}
	\item Discrete symmetries such as e.g. 
	\begin{enumerate}
		\item Parity symmetry
		\item Charge conjugation symmetry
		\item Time reversal symmetry
	\end{enumerate}

\noindent
do not lead to new conserved quantities in classical mechanics. They are important in quantum mechanics.
\end{itemize}

Symmetries tell us which interaction is responsible for particular process. Electromagnetic and strong interactions conserve parity and charge conjugation symmetry. However, it was observed that weak interactions violate these symmetries. This violation was first discovered in 1957 for parity symmetry \cite{Wu} and it was thought that a combination of C and P (a CP symmetry) is conserved. Yet, in 1964 it was discovered in experiment with kaons, that CP symmetry is also violated. In the decays of neutral kaons it occurs at a level of $10^{-3}$ \cite{Christenson}. Next assumption was that a combination of C, P and T (a CPT symmetry) is conserved. Until now nobody has discovered violation of this symmetry. In this thesis I concentrate on the CP symmetry therefore P and C symmetries will be described more detailed.

%%%%%%%%%%%%%%%%%%%%%%%%%%%%%

\subsection{Parity symmetry} \label{subsec:parity_symmetry} 

The contents of this and following subsection are mostly based on Ref. \cite{Martin}.

Parity is related to spatial reflection. Spatial reflection of polar vector $\vec{x}$ gives $-\vec{x}$. Thus the action of the parity operator $\hat{P}$ on a single-particle state represented by wave function $\psi(\vec{r},t)$ gives:

\begin{equation}
	\hat{P} \psi(\vec{x},t) = X\cdot \psi(-\vec{x},t),
	\label{s_ps_eq1}
\end{equation} 

\noindent	
where $X$ is a phase factor. If wave function $\psi(\vec{r},t)$ is eigenstate of parity operator then:

\begin{equation}
	\hat{P} \psi(\vec{x},t) = P\cdot \psi(\vec{x},t),
	\label{s_ps_eq01}
\end{equation} 

\noindent
where the eigenvalue $P=\pm 1$ is called the intrinsic parity or parity of the state. Parity is a multiplicative quantum number. 

If a particle has an orbital angular momentum $l$, there is an additional contribution to the total parity of the system. In this case particle's wave function consists of a radial part $R_{nl}(\vec{r})$ and an angular part $Y^{m}_{l}(\theta,\phi)$:

\begin{equation}
	\psi_{nlm}(\vec{x})=R_{nl}(\vec{r})\cdot Y^{m}_{l}(\theta,\phi),
	\label{s_ps_eq2}
\end{equation}

\noindent
where $n$ is principal and $m$ is magnetic quantum number. $Y^{m}_{l}(\theta,\phi)$ is a spherical harmonic. It can be shown that the parity transformation $\vec{x} \to -\vec{x}$ for spherical coordinates implies that:
	
\begin{equation}
	r \to r, \qquad \theta \to \pi-\theta, \qquad \phi \to \pi+\phi,
	\label{s_ps_eq3}
\end{equation} 

\noindent
and

\begin{equation}
	 Y^{m}_{l}(\theta,\phi) \to Y^{m}_{l}(\pi-\theta, \pi+\phi) = (-1)^{l}\cdot Y^{m}_{l}(\theta,\phi).
	 \label{s_ps_eq4}
\end{equation}

\noindent
Therefore,

\begin{equation}
	\hat{P} \psi_{nlm}(\vec{x}) = P\cdot \psi_{nlm}(-\vec{x}) = P\cdot(-1)^{l}\cdot \psi_{nlm}(\vec{x}),
	\label{s_ps_eq5}
\end{equation}

\noindent
where $\psi_{nlm}(\vec{x})$ is an eigenstate of parity operator $\hat{P}$ with eigenvalue $P\cdot(-1)^{l}$.  

Strong and electromagnetic interactions are invariant under parity. Parity transformation does not change the Hamiltonian of the system governed by strong and electromagnetic interactions. 
%%%%%%%%%%%%%%%%%%%%%%%%%%%%

\subsection{Charge conjugation symmetry} \label{subsec:charge_conjugation_symmetry}
% \cite{Martin}
% s 12

\indent 
Charge conjugation is the operation that changes particles into their antiparticles. The result of action of charge conjugation operator for particles which are identical to their antiparticles $(a \equiv \bar{a})$ is different from the result if particle differs from its antiparticle $(b \neq \bar{b})$:

\begin{equation}
\hat{C} \left| a,\psi_{a} \right> = C_{a} \left| a,\psi_{a} \right>
	\label{s_ccs_eq1}
\end{equation}

\begin{equation}
\hat{C} \left| b,\psi_{b} \right> = \left| \bar{b},\psi_{\bar{b}} \right>,
	\label{s_ccs_eq2}
\end{equation}

\noindent
where $C_{a}=\pm1$ is a phase factor, letter $a$ in ket denotes particle and $\psi_{a}$ corresponds to its wave function. Phase factor for neutral pion is equal to $+1$ $(C_{a}(\pi^{0})=+1$). 

Particles can be divided into two groups. The first group comprises particles which wave functions are eigenfunctions of $\hat{C}$ operator (\ref{s_ccs_eq1}). These particles are also theirs own antiparticles. Neutral pions, $\eta$ mesons and gamma quanta are examples of this kind of particles. In the second group particles and corresponding antiparticles are different (\ref{s_ccs_eq2}). Examples for this case are charged pions and protons.

Charge conjugation is multiplicative quantum number \cite{Perkins}, operator $\hat{C}$ acting on a system of two neutral pions $\psi\equiv\psi_{a}\psi_{a}$ gives:

\begin{equation}
\hat{C} \left| \psi \right> = C_{a} \left|a, \psi_{a} \right> C_{a} \left|a, \psi_{a} \right> = C_{a}^{2} \left| a, \psi_{a} \right> \left|a, \psi_{a} \right>.
	\label{s_ccs_eq3}
\end{equation}
%%%%%%%%%%%%%%%%%%%%%%%%%%%%%%%%%%%%%%%%%%%%%%%%%%%%%%%%%%%%%%%%%%%%%%%%%%%%%%%%%%%%%%%

\newpage

\begin{center}\end{center}
\newpage

%%%%%%%%%%%%%%%%%%%%%%%%%%%%%%%%%%%%%%%%%%%%%%%%%%%%%%%%%%%%%%%%%%%%%%%%%%%%%%%%%%%%%%%%%%
\newpage
\thispagestyle{plain}

\pagestyle{fancy}
\fancyhf{} 
\fancyhead[LE,RO]{\textbf{\thepage}}
\fancyhead[RE]{\small\textbf{{CP symmetry violation in the $\eta$ meson decay}}} %tytul rozdzialu na stronie parzystej

\newpage

\section{CP symmetry violation in the $\eta$ meson decay} \label{sec:CPV}

\vspace{0.5cm}

To understand why C and P symmetries are violated in the $\eta$ meson decays, it is necessary to know a wave function of the final state system. It is worth to remember that the total wave function of a final state is a product of spin, isospin and spatial parts:

\begin{equation}
\psi_{total}=\psi_{spin}\cdot\psi_{isospin}\cdot\psi_{spatial}.
	\label{cpvimed_eq1}
\end{equation}

\noindent 
Therefore, resultant sign of the wave function under the parity transformation is the product of parities of those three components. 

In the following subsections each part of the total wave function will be considered for two and four pion systems.

%%%%%%%%%%%%%%%%%%%%%%%%%%

\subsection{System of two neutral pions $(\eta \to \pi^{0}\pi^{0})$}

At the beginning of this subsection spin part will be considered. Spin of neutral pion is equal to zero, $S(\pi^{0})=0$. Therefore, the spin of a pair of pions is also equal to zero,  $S(\pi^{0}\pi^{0})=0$. In the spin space the parity operation is equivalent to the exchange of the spins in the wave function. Using the property of Clebsh-Gordan coefficients~\cite{Messiah2}: 

\begin{equation}
<s_{1}m_{1}s_{2}m_{2}|SM>=(-1)^{s_{1}+s_{2}-S}<s_{2}m_{2}s_{1}m_{1}|SM>,
	\label{c_ppp_eq1}
\end{equation}

\noindent
where $s_{i}$ denotes spin of i-th particle, $m_{i}$ is spin projection, $S$ stands for spin of $2\pi$ pair and $M$ corresponds to its projection, it can be shown that the sign of the spin part under the parity transformation of the wave function is positive ($\psi_{spin}=+1$). It is because $(-1)^{0+0-0}=+1$.

\indent
The isospin of a $\pi$ meson is equal to 1, $I(\pi)=1$. Hence, the isospin of two pions can be equal to $I_{2\pi}=0,1,2$. However, the Clebsh-Gordan coefficient for two neutral pions in the final state of isospin 1 equals to zero. That is why the isospin of ($\pi^{0}\pi^{0}$) system can be only $I_{2\pi^{0}}=0,2$. Using equation~(\ref{c_ppp_eq1}) for isospin, there are only two possibilities for $(-1)^{i_{1}+i_{2}-I}$ coefficient: $(-1)^{1+1-0}$ or $(-1)^{1+1-2}$; in both cases this factor is equal to~$+1$. 

\indent
Finally, as regards spatial part of the wave function based on equation~(\ref{s_ps_eq5}) it can be written that the parity of a wave function for two particles is the product of internal parities of each particle and the parity resulting from their orbital angular momentum~$L$: 

\begin{equation}
\hat{P}\psi(\equiv\psi_{1}\psi_{2})=P(\psi_{1})P(\psi_{2})(-1)^{L}.
	\label{c_ppp_eq3}
\end{equation}
 
The internal parity of pion is $P(\pi)=-1$. From equation~(\ref{c_ppp_eq3}) one can see that the change of sign of spatial part of $2\pi^{0}$ wave function under the parity transformation is equal to the factor $(-1)^{L}$. 

In order to evaluate value of $L$, information about the total angular momentum $\vec{J}$ of the decay $\eta \to \pi^{0}\pi^{0}$ can be used. Angular momentum is described by equation:

\begin{equation}
	\vec{J}=\vec{L}+\vec{S},
	\label{c_ppp_eq4}
\end{equation}

\noindent
where $\vec{L}$ denotes orbital angular momentum and $\vec{S}$ stands for the total spin of the system. Because a total angular momentum of initial and final state has to be equal and $J(\eta)=0$ so $J(\pi^{0}\pi^{0})$ also has to be equal to zero. Knowing that $S(\pi^{0}\pi^{0})=0$ and using equation (\ref{c_ppp_eq4}), one can get $0=\vec{L}+0$. Therefore, the orbital angular momentum of $\pi^{0}\pi^{0}$ system originating from the $\eta$ meson decay equals to zero, $L(\pi^{0}\pi^{0}) = 0$. Thus, the factor $(-1)^{L}$ in equation (\ref{c_ppp_eq3}) is equal to~$+1$.

Parity symmetry in $\eta \to \pi^{0}\pi^{0}$ decay can be determined by gathering above results. 
From the above considerations we may infer that the $\pi^{0}\pi^{0}$ system produced in $\eta \to \pi^{0}\pi^{0}$ decay possesses parity equal to $+1$ because the change of sign of spin, isospin and spatial parts of wave function under the parity transformation are all equal to $+1$ $(P(\psi_{spin})~=~+1,\ P(\psi_{isospin})~=~+1,$  $P(\psi_{spatial})~=~+1)$. Taking into account that $P(\eta) = -1$, one can see that $P$ symmetry is violated in $\eta \to \pi^{0}\pi^{0}$ decay. 

Charge conjugation symmetry phase factor $C_{a}$ of $2\pi^{0}$ system with $L=0$ has the same value as for $\eta$ and is equal to $+1$. Thus, $C$ symmetry in $\eta \to \pi^{0}\pi^{0}$ decay is conserved, but CP symmetry is violated because $P$ is violated.  

Consequently, in the $\eta \to \pi^{0}\pi^{0}$ decay a $C$ symmetry is conserved but $P$ and $CP$ symmetries are violated. 

%%%%%%%%%%%%%%%%%%%%%%%%%%%%%%%%%%%%

\subsection{System of four neutral pions $(\eta \to 4\pi^{0})$}

Equation (\ref{s_ccs_eq3}) implies that charge conjugation eigenvalue of the $4\pi^{0}$ system is just equal to $(+1)^{4}=+1$.

Spin and isospin parts of total wave function are analogous to those in the $2\pi^{0}$ system and they do not change its sign under the parity transformation. Spatial part of four particles system is more complex than for two and will be considered in detail in the following part. 

\begin{figure}[ht]
		\centering
		\includegraphics[width=0.35\textwidth]{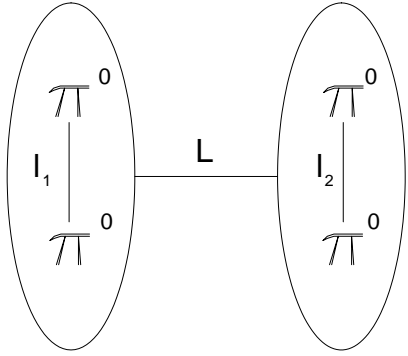}
			\caption{Picture illustrates angular momenta in four neutral pions system.}
			\label{rys_4piSystem}
	\end{figure}

System of $4\pi^{0}$ can be described, for example, in a way shown in Fig. \ref{rys_4piSystem}. In this case there are three orbital angular momenta $l_{1},l_{2},L$, which need to be considered.
	
Due to Bose symmetry $l_{1}$ and $l_{2}$ are even \cite{Perkins, Messiah2}. Thus, in general, there are three possibilities of orbital angular momenta in the system: 

\begin{enumerate}

\item Orbital angular momenta in two pion pairs are equal, $l_{1}=l_{2}$. Four neutral pions system is a bosonic system, so due to Bose symmetry $L$ is also even. Substituting it in (\ref{s_ps_eq5}), when all orbital angular momenta are even, one can calculate that parity of the system is equal to $P=+1$. Thus, in this case $CP(4\pi^{0})=+1$ while $CP(\eta) = -1$, and so the CP is violated

\item Orbital angular momentum of one pion pair is zero, $l_{1}=0$ and of the other is even, $l_{2}=2n$, where $n$ is a natural number. In this case $L$ is even because $J(\eta)=0$, so $J(4\pi^{0})$ also has to be zero. It is possible if and only if $L$ and $l_{2}$ are equal. In this case CP is also violated.

\item Orbital angular momentum of one pion pair is at least two, $l_{1}\geq 2$ and of the other is at least four, $l_{2} \geq 4$. 
Considering $l_{1}=2$ and $l_{2}= 4$, these two vectors can be combined and one of possible result is three, $l_{1,2}=3$. If $L$ is also equal to three ($L=3$) then the total angular momentum in the $\eta \to 4\pi^{0}$ may be conserved, however one obtains $P=-1$. It is because $(-1)^{2+4+3}=-1$. In this case $CP(\eta)=CP(4\pi^{0})=-1$ and CP symmetry is conserved.

To check how probable is this case estimation of orbital angular momentum values accessible in the system is needed. Excess of mass, which can be transformed into energy in $\eta \to 4\pi^{0}$ decay is approximately $8 \ MeV/c^{2}$. Angular momentum in classical mechanics is given by $L = \vec{r} \times \vec{p}$. In order to estimate a maximum angular momentum	accessible in a $4\pi^{0}$ system, one can make a conservative assumption that one pion pair do not take any energy. Assuming that remaining two pions share accessible energy equally, one can find $|\vec{p}|=68\ MeV/c$. For $\vec{r}$\ it can be assumed that radius of $\eta$ meson is not larger than radius of nucleon ($\sim 1 \ fm$). Substituting these values into equation for angular momentum we obtain $L=68	\ MeV\cdot fm/c$. The obtained  value of $L$ is much smaller compared to the unit of angular momentum $\hbar \approx 197 MeV\cdot fm/c$ \cite{PDG}. Since we need at least four units in one pion pair, $L$ is only $9\% $ of needed value. In addition, it is worth to remember that to the CP symmetry conservation also two units of orbital angular momentum in the second pair of pions are necessary. It is estimated that branching ratio for process with CP symmetry conservation in $\eta \to 4\pi^{0}$ decay is less than $10^{-10}$ \cite{Kupsc:2011ah}.

\end{enumerate}

%%%%%%%%%%%%%%%%%%%%%%%%%%%%%%%%%%%%%%%%%%%%%%%%%%%%%%%%%%%%%%%%%%%%%%%%%%%%%%%%%%%%%%%%%%%%%%%%
\newpage
\thispagestyle{plain}

\pagestyle{fancy}
\fancyhf{} 
\fancyhead[LE,RO]{\textbf{\thepage}}
\fancyhead[RE]{\small\textbf{{Reaction $pp \to pp\eta$ and $pp \to pp \ multipion$}}}

\newpage

\section{Reaction $pp \to pp\eta$ and $pp \to pp \ multipion$} \label{reactions}

\vspace{0.5cm}

\begin{figure}[h] 
	\centering
	\includegraphics[width=0.9\textwidth]{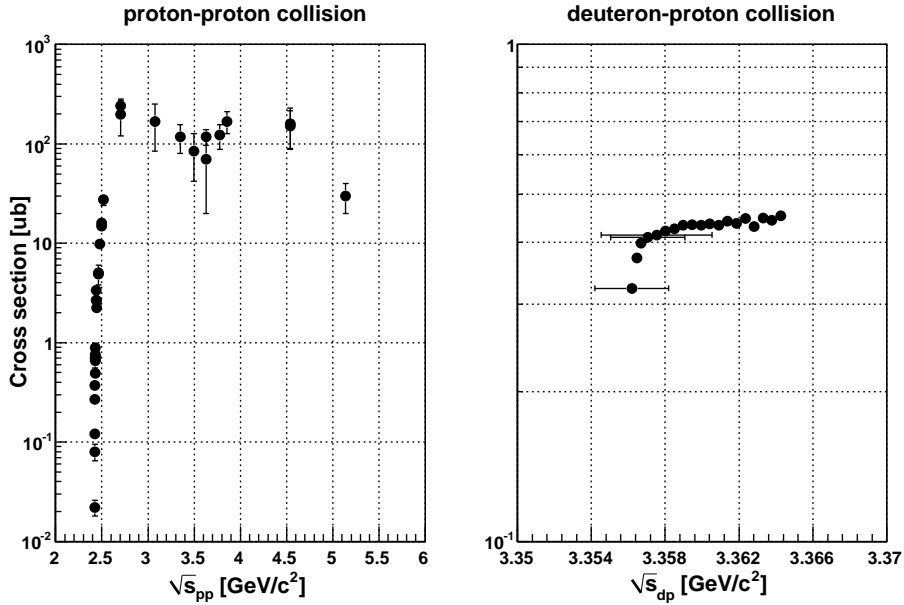}
		\caption{Cross sections for the $\eta$ meson production in proton-proton (left) and proton-deuteron collisions (right).  The left hand side histogram contains experimental data from Refs. \cite{Chiavassa, Pauly_art, Smyrski, Hibou, Flaminio:1984gr, Moskal:2003gt} and the right hand side histogram shows data from Ref.~\cite{Smyrski_hel, Mersmann:2007gw}. }
		\label{rys_cross2eta}
\end{figure}

The Cooler~Synchrotron in Forschungszentrum~J\"ulich (see subsection \ref{subsec:COSY}) enables $\eta$ production in two main reactions: $pp \to pp\eta$ and $dp \to \  
^3\!He \hspace{0.11cm} \eta$. Cross sections as a function of $\sqrt{s}$ for these reactions are shown in Fig. \ref{rys_cross2eta}. 

One can see that the shape of the cross sections as a function of $\sqrt{s}$ is quite similar. Starting close to threshold of the reaction they can be described by rapid increase in cross section value and later transfers into plateau. The plateau for proton-proton collision is at a level of $10^2\ \mu b$ and for deuteron-proton collision it is located at above $0.4\ \mu b$. Since the $\eta \to 4\pi^{0}$ is a rare decay, a high statistics of $\eta$ production is required. Thus, proton-proton collision has been chosen and will be considered in further analysis.

For the estimation of the precision for the extraction of a BR of $\eta \to 4\pi^{0}$ decays we need to estimate also the physical background. The background is related to techniques which are used for decay particles reconstruction. 

At the \mbox{WASA-at-COSY} experiment the missing mass and invariant mass techniques are used for $\eta$ meson identification via the $pp\to ppX$ reaction where X denotes the unobserved particle. If protons four-momenta $\mathbb{P}\equiv(E,\vec{p})$ before and after the collision can be determined, then conservation of a total four-momentum vector gives:

\begin{equation}
\mathbb{P}_{beam} + \mathbb{P}_{target} = \mathbb{P}_{1} + \mathbb{P}_{2} + \mathbb{P}_{X},
\label{rppte_eq4}
\end{equation}

\noindent
where $\mathbb{P}_{X}$ stands for the four-momentum of unobserved particle. Employing equation (\ref{rppte_eq4}) the mass $m_{X}$ can be calculated as:

\begin{equation}
m_{X}^{2}=|\mathbb{P}_{X}|^{2}=|\mathbb{P}_{beam} + \mathbb{P}_{target} - \mathbb{P}_{1} - \mathbb{P}_{2}|^{2}.
\label{rppte_eq5}
\end{equation}

Another useful method of unstable particle identification is invariant mass technique. It is based on a reconstruction of a rest mass of decaying particle which can be calculated if information about all decayed products is known. The mass is described by:

\begin{equation}
m = \sqrt{ \Big(\sum_{i}{E_{i}} \Big)^{2} - \Big|\sum_{i}{\vec{p}_{i}} \Big|^{2} },
\label{rppte_eq6}
\end{equation}

\noindent
where $E_{i}$ and $\vec{p}_{i}$ denote energy and momentum of i-th particle respectively.

In spite of using both techniques the $\eta \to 4\pi^{0}$ decay can be misidentified due to the unavoidable physical background as well as due to the misinterpretation of the detector signals. The main physical background of $\eta \to 4\pi^{0}$ is prompt $4\pi^{0}$ production in proton-proton collision. This reaction is difficult to reject because all particles in exit channel for both reactions are the same. Difference between mass of $\eta$ and $4\pi^{0}$ is only approximately $8\ MeV/c^{2}$. Resolution of $\mbox{WASA-at-COSY}$ detector is not good enough to reduce the background events using invariant mass technique.

\begin{figure}[h] 
	\centering
	\includegraphics[width=0.8\textwidth]{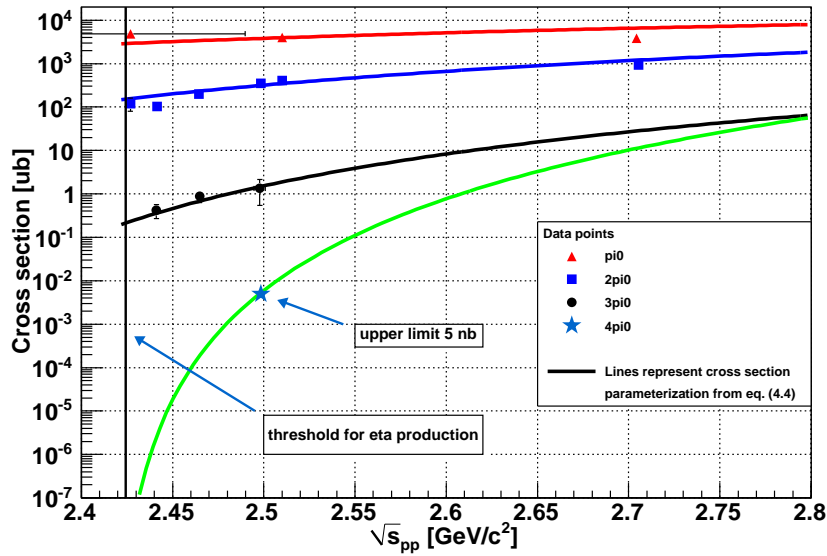}
		\caption{Experimental cross section for direct pions production in proton-proton collision (points). Superimposed lines denotes result of the fit of formula \ref{rppte_eq1} to the experimental points. }
		\label{rys_przekrojPiony}
\end{figure}

The current available data of cross section for multipion production in proton-proton collision are shown in Fig. \ref{rys_przekrojPiony}. The cross section in the figure is shown as a function of excess energy with respect to the $\eta$ meson production threshold.

The cross section for $4\pi^{0}$ production in proton-proton collisions is not established and so far only an upper limit for a single energy point at $\sqrt{s} = 2.498\ GeV/c^{2}$ was determined to $5\ nb$ \cite{Pauly_phd}. In order to estimate the background from direct pions production $(pp \to pp4\pi^{0})$ for the $pp \to pp\eta \to pp4\pi^{0}$ reaction we have derived an energy dependence of the upper limit of the total cross section under assumption of the homogeneous phase-space population \cite{Byckling}:

\begin{equation}
	\sigma(Q) = const \cdot (Q/\sqrt{s})^{(3m-5)/2},
	\label{rppte_eq1}
\end{equation}

\noindent 
where $Q$ and $s$ denote the excess energy and the total energy in the centre-of-mass system respectively, $m$ stands for the number of particles in the exit channel.

The result is shown in Fig. \ref{rys_przekrojPiony} as the lowest, green line for which the absolute value was fitted to the experimental upper limit. For the comparison in Fig. \ref{rys_przekrojPiony} also cross sections for $\pi$ , $2\pi$ and $3\pi$ production are shown. The superimposed lines indicate result of the fit of formula \ref{rppte_eq1} with amplitude as the only free parameter. The obtained values of $const$ are listed in Tab.~\ref{normalization_tab}. 

\begin{center}
\begin{table}[h!]
	\centering
	\begin{tabular}{c|c}
		\hline 
		Number of neutral pions &  Normalization $const$ from equation \ref{rppte_eq1}  \\ 
		\hline
		$1$ & $10^{5}$  \\
		$2$ & $3\cdot10^{5}$  \\
	  $3$ & $3\cdot10^{5}$  \\
	  $4$ & $2.35\cdot10^{7}$  \\
	\end{tabular}\\
	\caption{Values of $const$ parameter determined by fitting parameterization \ref{rppte_eq1} to experimental data shown in Fig.~\ref{rys_przekrojPiony} }
	\label{normalization_tab}
\end{table}
\end{center}

In Fig. \ref{rys_przekrojPiony} one can see that e.g. at 50 MeV above the $\eta$ meson production threshold the cross section for $4\pi$ production is by about five orders of magnitude smaller than for $2\pi$ production. Consequently, the physical background for studies of $\eta \to 2\pi^{0}$ decay is by five orders of magnitude larger than for $\eta \to 4\pi^{0}$ decay. Moreover, with decreasing excess energy this ratio is increasing.

%%%%%%%%%%%%%%%%%%%%%%%%%%%%%%%%%%%%%%%%%%%%%%%%%%%%%%%%%%%%%%%%%%%%%%%%%%%%%%%%%%%%%%%%%%%%%%%%%
\newpage
\thispagestyle{plain}

\fancyhf{} 
\fancyhead[LE,RO]{\textbf{\thepage}}
\fancyhead[RE]{\small\textbf{{Experimental facility}}} 

\newpage

\section{Experimental facility} \label{sec:WASA}

\vspace{0.5cm}

\subsection{Cooler Synchrotron COSY} \label{subsec:COSY}

The contents of this and the following subsection are mostly based on Refs.~\cite{Adam, Redmer, Koch}.

The COoler SYnchrotron (COSY) is located at the Institute of Nuclear Physics of the Research Centre J\"ulich in Germany. It is a storage ring of 184 m circumference. The facility provides unpolarized and polarized proton and deuteron beams in the momentum range from $0.3 \ GeV/c$ up to $3.7 \ GeV/c$. To decrease the momentum and spatial spread of the COSY beam, electron and stochastic cooling is used.    

\subsection{WASA detector}
\begin{figure}[ht] 
		\centering
		\includegraphics[angle=-90, width=0.8\textwidth]{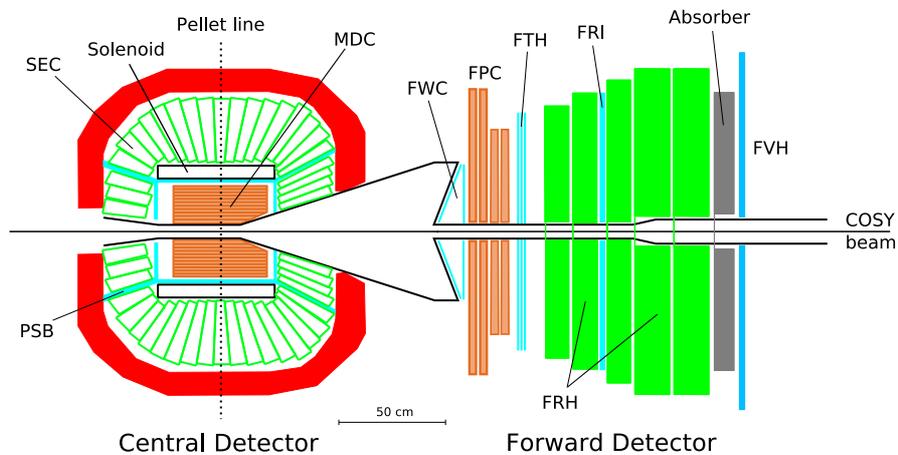}
			\caption{Schematic view of the WASA-at-COSY detector. The abbreviations of the name of detectors are explained in the text. }
			\label{rys_wasa_det}
	\end{figure}

The Wide Angle Shower Apparatus (WASA) is approximately $4\pi$ angle detector and is located at the COSY ring. It consists of two main parts: The Forward Detector (FD) and The Central Detector (CD) as it is schematically shown in Fig.~\ref{rys_wasa_det} \cite{wasaWiki}.

\subsubsection{The Forward Detector}

The Forward Detector measures charged target-recoil particles and scattered projectiles. It covers scattering angles between $3^{\circ}$ and $18^{\circ}$, and consists of 30 layers of detectors:

\textbf{The Forward Window Counters (FWC)}\\
The FWC is the first detector in the FD along the beam direction. It consists of two layers and 48 elements altogether. Each element is made of plastic scintilator with thickness of $3 \ mm$. The FWC is used as a trigger. Coincident hits in different subdetectors at the same azimuthal angle are used to select events originating from the interaction vertex.

\textbf{The Forward Proportional Chambers (FPC)}\\
The FPC is mounted directly behind the FWC tracking device downstream the beam. It consists of four modules, each with four layers of 122 proportional drift tubes (straws). The modules are rotated by $45^{\circ}$ with respect to each other (in the plane perpendicular to the beam axis). The FPC is used to reconstruct tracks of charged particles.

\textbf{The Forward Trigger Hodoscope (FTH)}\\
The FTH is third sub-detector. It is made of three layers built out of $5 \ mm$ plastic scintilators. The first layer consists of  48 radial elements with shape similar to triangles. Second and third layers are divided into 24 elements shaped as Archimedian spirals. It provides hits multiplicity and both angles: polar and azimuthal. The FTH is mainly used in the first level trigger logic in a coincidence with other sub-detectors.

\textbf{The Forward Range Hodoscope (FRH) }\\
The FRH consists of 5 layers made of plastic scintilator, each layer is comprised of 24 elements. First three layers have thickness of $110 \ mm$, while thickness of fourth and fifth layer is equal to $150 \ mm$. The FRH is used for kinetic energy reconstruction and particle identification.

\textbf{The Forward Range Intermediate Hodoscope (FRI)}\\
The FRI is two layered detector placed between the second and the third layer of the Forward Range Hodoscope. Each of its layer is made of 32 rectangular elements. Bars in one layer are oriented horizontally and vertically in the other. The purpose of placing FRI was to improve two-dimensional position sensitivity within the FRH.

\textbf{The Forward Veto Hodoscope (FVH)}\\
The last active element of the Forward Detector is FVH. It is made of 12 horizontal plastic scintilator bars with thickness of $2 \ cm$. It is used for detecting high energetic particles. Lately second layer of FVH was installed. It will be used as a stop detector for Time-of-Flight measurements. This will improve the energy resolution for high energy particles \cite{Zielinski}.

\textbf{The Forward Range Absorber (FRA)}\\
The FRA is a passive absorber layer made from iron plates with thickness of $5 \ mm$. Maximum thickness of the absorber is $100 \ mm$.  It can be installed between the FRH and FVH detectors.  For adequate thickness of the absorber slower protons from reactions in which $\eta$ is produced are stopped in the absorber, while faster protons and pions ($\eta$ is not produced) can give signals in FVH. 

\subsubsection{The Central Detector}

The Central Detector (CD) surrounds the interaction point. It consist of the Mini Drift Chamber, the Plastic Scintillator Barrel, the Superconducting Solenoid, the Scintillating Electromagnetic Calorimeter. All these components allow for measurement of momentum and energy of neutral and charged particles.

\textbf{The Mini Drift Chamber (MDC)}\\
The MDC is placed around the beam pipe. It is assembled from 17 cylindrical layers with 1738 straw tubes. It covers scattering angles between $24^{\circ}$ and $159^{\circ}$. The MDC is used for the
determination of the charged particle momenta and the interaction vertex.

\textbf{The Plastic Scintillator Barrel (PSB)}\\
The PSB is a cylindrical detector surrounding the drift chamber. It consists of a central barrel part (cylindrical) and two end caps (forward and backward). Caps are made of trapezoidally shaped 8 mm thick plastic scintillator elements. Cylindrical part consists of 2 layers composed of rectangular strips with a small overlap between neighbouring elements. The PSB is used to distinguish charged form neutral particles and serves as $\Delta E$ detector for charged particles identification.

\textbf{The SuperConducting Solenoid (SCS)}\\
Magnetic field is necessary to determine momenta of charged particles with the MDC. It can be done by dint of the SCS. The solenoid surrounds the Mini Drift Chamber and the Plastic Scintillator Barrel. It provides axial magnetic field used for bending charged particle trajectories. Cooling of the solenoid is provided by liquid Helium. The maximum axial magnetic field is up to $1.3 \ T$ in the interaction region.

\begin{figure}[h] 
		\centering
		\includegraphics[width=0.4\textwidth]{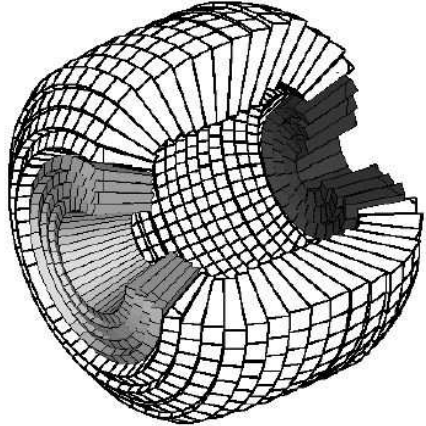}
			\caption{The Scintillating Electromagnetic Calorimeter (SEC) used in detector WASA-at-COSY. It is divided into three parts, marked in different colours,  starting from left: forward, central and backward part. }
			\label{rys_wasa_sec}
\end{figure}

\textbf{The Scintillating Electromagnetic Calorimeter (SEC)}\\
The SEC enables measurement of scattering angles as well as energy of gamma quanta, electrons and positrons. Energy of gamma quantum can be measured in the range from $2\ MeV$ up to $800\ MeV$. The calorimeter consists of 1012 sodium-doped CsI scintillating crystals grouped into three parts shown in Fig. \ref{rys_wasa_sec} with different colours.  Starting from left: forward, central and backward part. Calorimeter covers full azimuthal angle ($\phi$) and polar angle ($\theta$) from  $20^{\circ}$ to $169^{\circ}$ which corresponds to $96\%$ of full solid angle. The crystals are shaped as truncated pyramids to conform spherical geometry of the calorimeter. Three parts of the calorimeter differ in polar angle range and crystal size. The lengths of the crystals vary from 30 cm (16.2 radiation lengths) to 20 cm. The longest are in the forward part, the shortest in the backward part. The forward part consists of 4 layers with 34 elements each covering polar angle from $20^{\circ}$ to $36^{\circ}$. In the central part $\theta$ is between $34^{\circ}$ and $150^{\circ}$. It consists of 17 layers and 48 elements in each layer. The backward part covers polar angle from $150^{\circ}$ to $169^{\circ}$ and is built out of 3 layers. Two of them has 24 elements and one layer possesses only 12 elements.

Size of crystals depend on part in which they are placed and are crucial for angular resolution of the calorimeter (see subsection \ref{subsec:cluster_routine}). The energy resolution for gamma quanta is described as $\frac{\Delta E}{E} = \frac{5 \%}{\sqrt{E/GeV}}$. For stopped charged particles the energy resolution is equal to about $ 3\%$.

Each crystal is read out individually by photomultiplier. Readout is done outside of the iron yoke in order to avoid influence of magnetic field. 

More details on the construction and the design of the calorimeter can be found in Ref. \cite{Koch}.

%%%%%%%%%%%%%%%%%%%%%%%%%%%%%%%%%%%%%%%%%%%%%%%%%%%%%%%%%%%%%%%%%%%%%%%%%%%%%%%%%%%%%%%%%%%
\newpage
\thispagestyle{plain}

\fancyhf{} 
\fancyhead[LE,RO]{\textbf{\thepage}}
\fancyhead[RE]{\small\textbf{{Identification of $\eta \to 4\pi^{0}$ decay}}}

\newpage

\section{Identification of $\eta \to 4\pi^{0}$ decay} \label{sec:kinematics}

\vspace{0.5cm}

\subsection{Geometrical acceptance of calorimeter}
The first step of reaction $pp \to pp\eta \to pp4\pi^{0} \to pp8\gamma$ analysis is to check what fraction of all reactions can be fully observed with the WASA detector. For this purpose 1 000 000 reactions of $\eta$ meson production and decay into four neutral pions with the homogeneous phase-space distribution were generated by means of PLUTO program \cite{PLUTO}. Output file from the PLUTO was used in the WASA Monte Carlo (WMC) detector simulation. The WMC is based on the GEANT3 (GEometry ANd Tracking) software package developed at CERN \cite{CERNlib}. The GEANT program describes the passage of particles through the matter and simulates response of \mbox{WASA-at-COSY} detector. The analysis of detector signals is carried out using RootSorter. The RootSorter is a software package based on ROOT \cite{ROOT}.  

	\begin{figure}[h!] 
		\centering
		\includegraphics[width=0.75\textwidth]{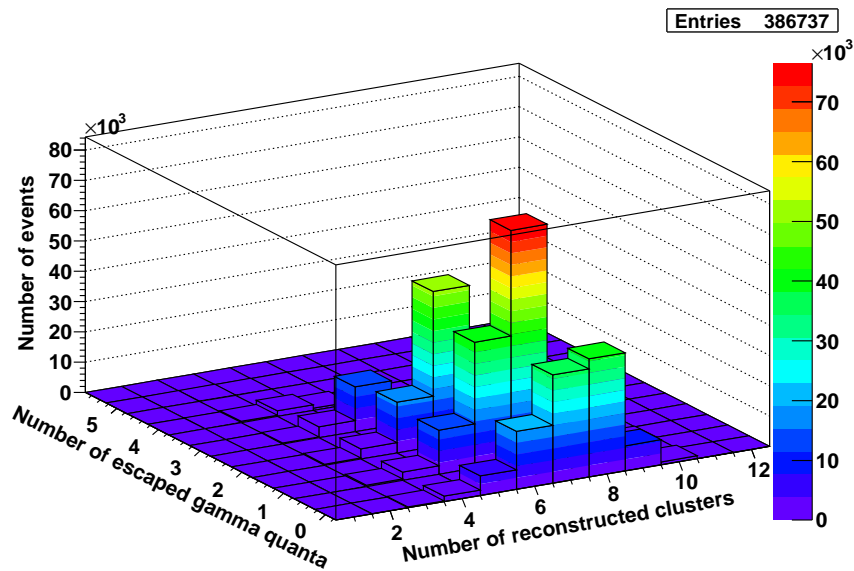}
			\caption{The number of reconstructed clusters for $pp \to pp\eta \to pp4\pi^{0} \to pp8\gamma$ reactions as a function of the number of gamma quanta which escaped from the Scintillating Electromagnetic Calorimeter. Histogram is drawn under the condition that two charged particles are reconstructed in the Forward Detector and there is no charged particle in the Central Detector.}
			\label{rys_numClustersVsOut}
	\end{figure}
 
The result of that studies is shown in Fig. \ref{rys_numClustersVsOut}. The figure presents the number of reconstructed clusters as a function of the number of gamma quanta which escaped from the Electromagnetic Calorimeter. A gamma quantum can escape if it flies into the beam pipe or the pellet pipe. Histogram is drawn under the condition that two charged particles were reconstructed in the Forward Detector and no charged particle in the Central Detector. These conditions are imposed in order to choose reactions when two protons fly into the FD and to reduce events when electron-positron pairs are created in the CD material. Reconstructed signal in SEC is considered to be a charged particle if corresponding  signal in PSB can be found. Only $38.7\%$ of all simulated events obey the above conditions.

\begin{figure}[h] 
		\centering
		\includegraphics[width=0.7\textwidth]{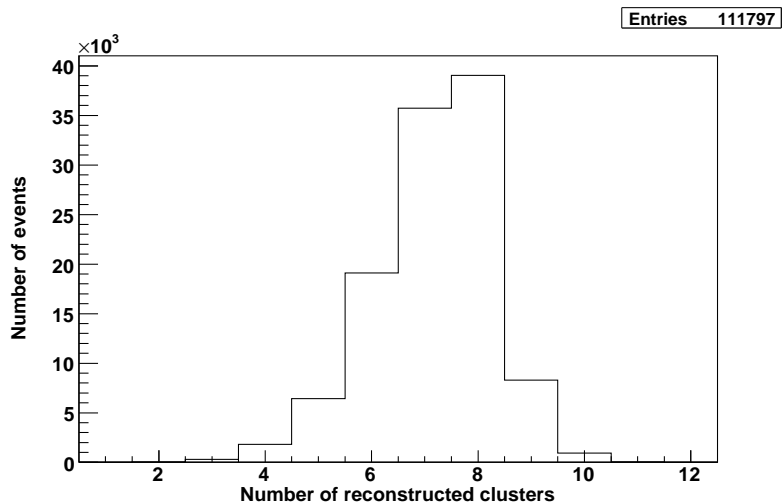}
			\caption{Projection of Fig. \ref{rys_numClustersVsOut} on X-axis for events when all eight gamma quanta from reaction $pp \to pp\eta \to pp4\pi^{0} \to pp8\gamma$ hit the Scintillating Electromagnetic Calorimeter.}
			\label{rys_numClusters}
	\end{figure}

When one or more gamma quanta escape from the SEC the information which they carry is lost. Therefore, for further consideration, only events when all eight gamma quanta hit the SEC are taken into account. Distribution of reconstructed clusters for the situation when all eight gamma quanta hit the SEC is shown in Fig. \ref{rys_numClusters}. It reduces number of accepted events to 11.2\%.

The histogram \ref{rys_numClusters} is plotted for events where no gamma quanta escaped from the SEC. One can see that the number of identified cluster varies from 3 to 10. This is a manifestation of merging and splitting of detected signals which leads to wrong clusters reconstruction (see subsection \ref{subsec:cluster_routine}).

\subsection{Pions reconstruction} \label{subsec:pions_reco}

For reaction where eight gamma quanta are in exit channel, in the further step of the analysis it must be established which gamma quanta originate from a decay of the same pion. That is why a routine which matches each two gamma quanta into pairs was elaborated. Purpose of the routine is to identify pairs of gamma quanta originating from the decay of the same pion. 
The prepared procedure first calculates squared invariant mass of every possible pair of gamma quanta and subtracts the squared mass of $\pi^{0}$. Next the results are combined into sets of four pairs which includes all eight gamma quanta. Further on absolute values of the differences between the invariant mass of the pair and $\pi^{0}$ mass within a given set are added together. The set for which the sum of differences from the pion mass is the smallest is assumed as a correct assignment and is taken for the further analysis.

If a number of reconstructed clusters is less than eight, there is not enough information to reconstruct the reaction $pp \to pp\eta \to pp4\pi^{0} \to pp8\gamma$. That is why reconstruction of the $\eta$ meson decay is done only for events when eight or more clusters are observed which reduces the number of accepted events further from $11.2\%$ to $4.8\%$.

\begin{figure}[h]
	\centering
	\includegraphics[width=0.85\textwidth]{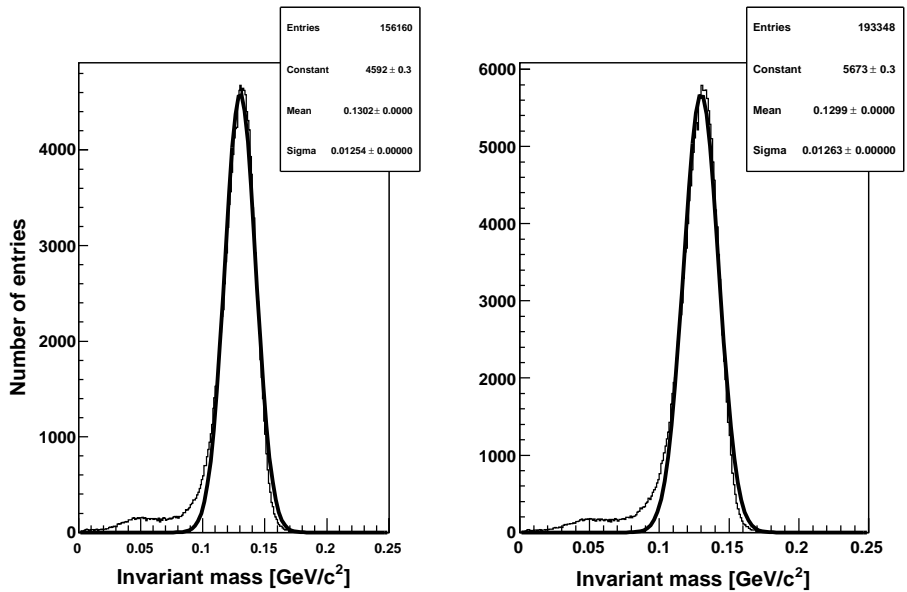}
	\caption{Invariant mass of pairs of gamma quanta originating from the $\eta$ meson decay, selected by means of algorithm described in the text. 
	The left-side histogram corresponds to events with eight reconstructed clusters, the right-side histogram contains events with eight or more reconstructed clusters. Gaussian functions (\ref{gaus}), fitted to the histograms, are shown as solid line.}
	\label{rys_invMpi0New}
\end{figure}

Fig. \ref{rys_invMpi0New} shows invariant mass of pairs of gamma quanta identified by above described algorithm.
Left panel presents the result for events with eight clusters reconstructed and the right histogram shows events when eight and more clusters have been reconstructed. To both histograms Gaussian function is fitted: 

\begin{equation}
Constant\cdot exp \Big(\frac{-(x-Mean)^{2}}{2\cdot Sigma^{2}} \Big).
\label{gaus}
\end{equation}

\noindent
Values of fit parameters are included inside the figures. Mean and sigma values do not differ significantly. However, statistics is larger for events with eight and more clusters reconstructed. Therefore, these events will be considered during further analysis.

Invariant mass spectra in Fig. \ref{rys_invMpi0New} are not symmetric and one can see tails at the left sides of both histograms. The tail is related to the functioning of two procedures. The first one matches gamma quanta into pairs and is described above. The second is algorithm which bases on energy deposited in the SEC and changes it into clusters information. To understand this process, knowledge of the cluster building routine is needed. 

\subsection{Cluster building routine}  \label{subsec:cluster_routine}

Description of the cluster building routine is mostly based on Refs. \cite{Koch, Hejny}.

Gamma quanta give signals in the SEC crystals by electromagnetic showers production. Charge of the signal provides information about energy deposited in a given detection module. 
Every crystal has its own readout system so it is taken as a separate element. To reconstruct observed particles, the routine combines all available hits into clusters. 

\begin{figure}[h]
	\centering
	\includegraphics[width=0.30\textwidth]{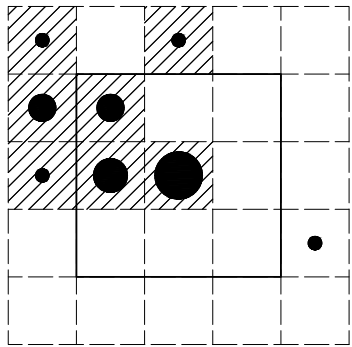}
	\caption{Ilustration of cluster creation.}
	\label{rys_algorytm}
\end{figure}

Cluster finding routine is an iterative procedure. First, it searches for the module with the highest energy deposition. The module becomes a reference element for a new cluster. Later it creates a square of $3\times3$ elements with the reference element in the centre. If any element with deposited energy is inside the square, it will be added to the cluster. All new added elements become also reference elements in next step of procedure. The procedure stops reconstruction of a cluster when no new elements with an energy deposition higher than $2\ MeV$ in the square are found. Further on the routine repeats the cluster finding excluding elements assigned to the already identified clusters. The procedure is repeated until all modules with deposited energy are attached to clusters. An example of cluster with structure of hit crystals is shown in Fig. \ref{rys_algorytm}. In the shown example elements with lined background are parts of a reconstructed cluster. One element, at right side is not attached to the cluster.

Restrictions for neighbouring elements which can be added in a cluster are included in the routine. Default maximal time difference between elements must be smaller than $50\ ns$. Minimal energy deposits are $5\ MeV$ for a new cluster starting element and $2\ MeV$ for a neighbouring element. Total energy deposited in a cluster is a sum of energy deposition in all crystals from which the cluster is built. Total energy of a cluster must be larger than $10\ MeV$. If it is smaller than $10\ MeV$, the cluster is rejected from further analysis.

The cluster time and cluster coordinates are calculated as a mean value of all attached elements weighted by the energy deposition. Energy deposited in a cluster is used to calculate gamma quantum kinetic energy. Particle  identification as a neutral requires no corresponding signal in the PSB. 

\begin{figure}[h]
	\centering
	\includegraphics[width=0.95\textwidth]{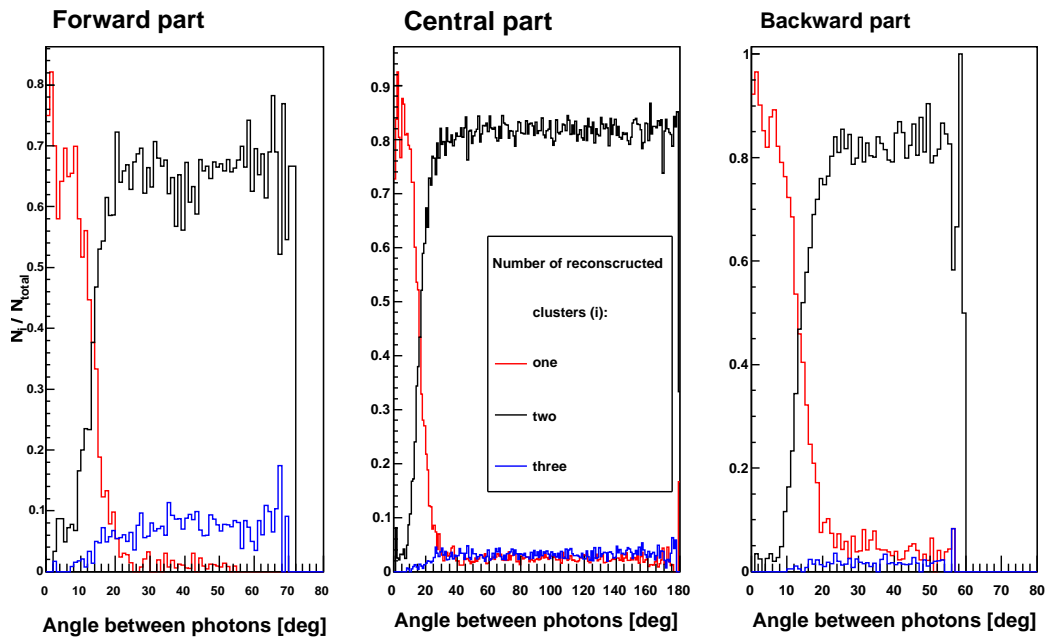}
	\caption{Probability for reconstruction of a given number of clusters for two gamma quanta hitting the calorimeter as a function of opening angle between them.}
	\label{rys_probability}
\end{figure}

Test of the ability of the used algorithm to reconstruct clusters depending on the opening angle between two gamma quanta was performed. Two gamma quanta with different opening angle were simulated in the test. Results are shown in Fig. \ref{rys_probability}. If opening angle between two gamma quanta is small, they are recognize as one cluster. It is enough that there is no 'empty' crystals between elements with non-zero energy deposition information. One cluster reconstruction dominate for small value of the opening angle (red line). Value of opening angle for which two cluster recognition is starting to be dominant depends on part of the calorimeter. It is connected with different size of crystals in different part of the SEC.

\begin{figure}[h]
	\centering
	\includegraphics[width=0.70\textwidth]{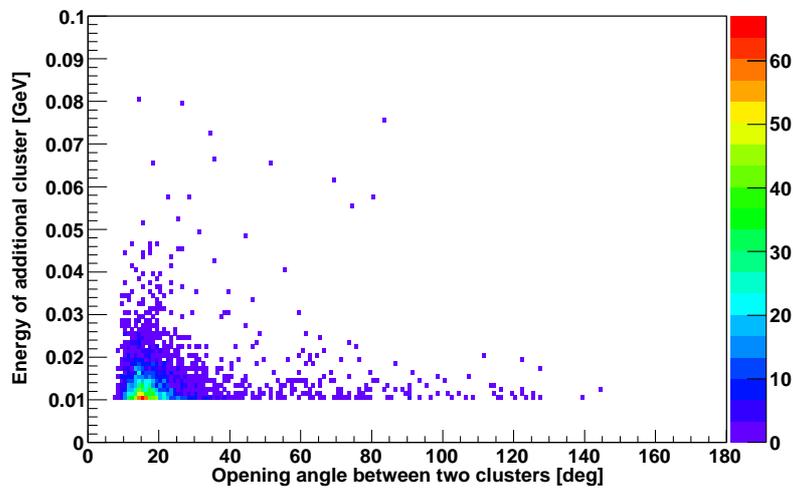}
	\caption{Histogram of energy deposited in an additional cluster (splitoff) as a function of opening angle between cluster generated by gamma quantum and the additional cluster.}
	\label{rys_fakeCluster}
\end{figure} 

In Fig. \ref{rys_probability}, apart from merged signals reconstructed as a single cluster (red line) one can see events when three clusters were recognized (blue line). Reconstruction of three clusters results from splitting of signals. One of possible reason for splitting is that electromagnetic shower caused by gamma quantum may spread across few modules. The middle crystal can be omitted in case if deposited energy is less than $2\ MeV$ or gamma quantum do not react in it at all. Thus, module read by the routine as 'empty' will be produced. To check this hypothesis, energy deposited in the additional cluster (splitoff) as a function of opening angle between cluster generated by gamma quantum and additional cluster is shown in Fig. \ref{rys_fakeCluster}. No entries below $10\ MeV$ are due to the requirement that the total energy of every cluster must be larger than $10\ MeV$. Distribution in Fig. \ref{rys_fakeCluster} shows that splitoffs are characterized by small energy deposition and small distance to real particle. But there is a small fraction of events for which the split cluster is far from the hit position of the gamma quantum.

\subsection{$\eta$ meson reconstruction} \label{subsec:eta_reconstruction}

At this point, when we already know how cluster building routine is working, we can go back to explain tails in the histograms shown in Fig. \ref{rys_invMpi0New}. As it is mentioned above, tails in histograms (\ref{rys_invMpi0New}) are artefacts of used algorithm and the granularity of the calorimeter.

\begin{figure}[h!]
	\centering
	\includegraphics[width=0.7\textwidth]{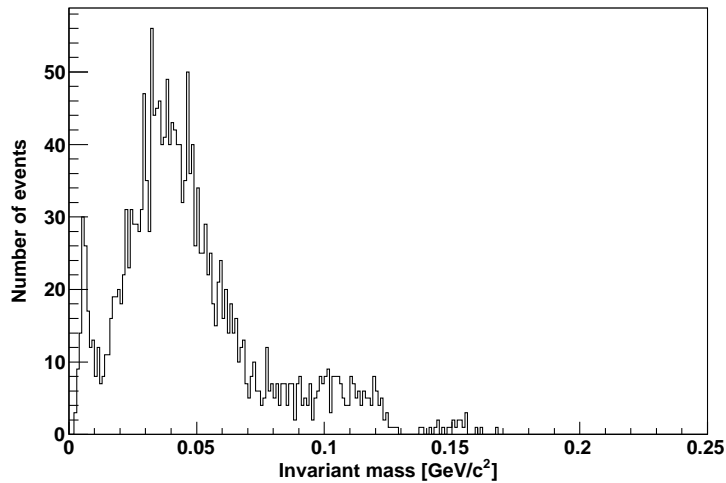}
	\caption{Histogram of invariant mass of cluster generated by gamma quantum and an additional one (splitoff).}
	\label{rys_invSplitoff}
\end{figure}

It may happen that in a given event number of merged signals is equal to the number of split ones. In this case one will reconstruct eight clusters as expected for the $pp \to pp\eta \to pp4\pi^{0} \to pp8\gamma$ reaction. Such event will fulfil all required criteria but nevertheless it will lead to the wrong reconstruction of energy and momenta of some gamma quanta. It manifests itself as a tail in the invariant mass histogram. In order to demonstrate this effect in Fig. \ref{rys_invSplitoff} invariant mass for pairs composed of real cluster and the additional one are plotted.

To held as little events with splitoffs as possible, only these pairs whose invariant mass differs from the pion mass by no more than $2\sigma$ are used in the further analysis. 

\begin{figure}[h!]
	\centering
	\includegraphics[width=0.7\textwidth]{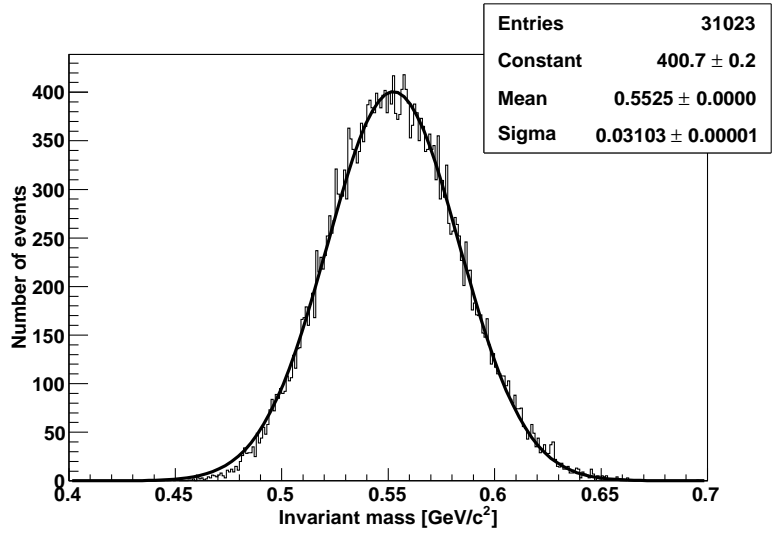}
	\caption{Invariant mass of eight gamma quanta from the $\eta$ meson decay. Solid line correspond to Gaussian function fitted to the histogram.}
	\label{rys_invMeta}
\end{figure}

Events in which all four gamma quantum pairs fulfil restrictions mentioned above are used for the $\eta$ meson reconstruction. Invariant mass of all eight gamma quanta is plotted in Fig. \ref{rys_invMeta}. 

\noindent
By fitting the Gaussian function to the histogram, a resolution of $\eta$ meson invariant mass reconstruction in $pp \to pp\eta \to pp4\pi^{0} \to pp8\gamma$ reaction with \mbox{WASA-at-COSY} detector is determined to $\sigma =31\ MeV/c^{2}$. 

Present software setup and restrictions imposed during analysis allow to reconstruct $\eta$ meson only for $3.1\%$ of all $pp \to pp\eta \to pp4\pi^{0} \to pp8\gamma$ events. 

%%%%%%%%%%%%%%%%%%%%%%%%%%%%%%%%%%%%%%%%%%%%%%%%%%%%%%%%%%%%%%%%%%%%%%%%%%%%%%%%%%%%%%%%%%%%%%%%
\newpage
\thispagestyle{plain}

\fancyhf{} 
\fancyhead[LE,RO]{\textbf{\thepage}}
\fancyhead[RE]{\small\textbf{{Time of measurement}}}

\newpage
\section{Time of measurement} \label{sec:time}

\vspace{0.5cm}

The main goal of this thesis is to estimate time of measurement, with the \mbox{WASA-at-COSY} detector, for the reaction $pp \to pp\eta \to pp4\pi^{0} \to pp8\gamma$ allowing to improve the present knowledge about the $BR(\eta\to 4\pi^{0})$. Up to now only an upper limit of the branching ratio has been known \cite{PDG}:

\begin{equation}
BR(\eta \to 4\pi^{0}) < 6.9\cdot 10^{-7}  \qquad CL=90\%.
\label{tom_eq1}
\end{equation}

\noindent
which is determined by Crystal Ball Collaboration \cite{CrystalBall}.

In general a number of reconstructed $\eta \to 4\pi^{0}$ decays can be written as

\begin{equation}
N_{\eta \to 4\pi^{0}}=BR(\eta\to4\pi^{0})\cdot A_{\eta} \cdot \sigma_{\eta} \cdot L \cdot \Delta t,
\label{tom_eq2}
\end{equation}

\noindent
where $N_{\eta \to 4\pi^{0}}$ and $BR$ stands for number of observed events and branching ratio respectively, $A_{\eta}$ is acceptance of the detector, $L$ denotes luminosity, $\sigma_{\eta}$ corresponds to cross section for $\eta$ meson production and $\Delta t$ is time of measurement.

Rewriting it as an equation for branching ratio, we find

\begin{equation}
BR(\eta\to4\pi^{0})=\frac{N_{\eta \to 4\pi^{0}}}{A_{\eta} \cdot \sigma_{\eta} \cdot L \cdot \Delta t}.
\label{tom_eq3}
\end{equation}

Statistical errors of branching ratio depends on uncertainty of all variables \\ ($\sigma(N_{\eta \to 4\pi^{0}}),\ \sigma(A_{\eta}),\ \sigma(\sigma_{\eta}),\ \sigma(L)$ and $\sigma(\Delta t)$). However, uncertainty related to $A_{\eta},\ \sigma_{\eta},$ \\ $L$ and $\Delta t$ can be assumed to be negligibly small with comparison to $\sigma(N_{\eta \to 4\pi^{0}})$. Under this assumption $\sigma(BR)$ may be expressed as:

\begin{equation}
\sigma(BR) = \Big| \frac{\sigma(N_{\eta \to 4\pi^{0}})}{A_{\eta} \cdot \sigma_{\eta} \cdot L \cdot \Delta t} \Big| 
\label{tom_eq5}
\end{equation} 

\begin{figure}[h]
	\centering
	\includegraphics[width=0.70\textwidth]{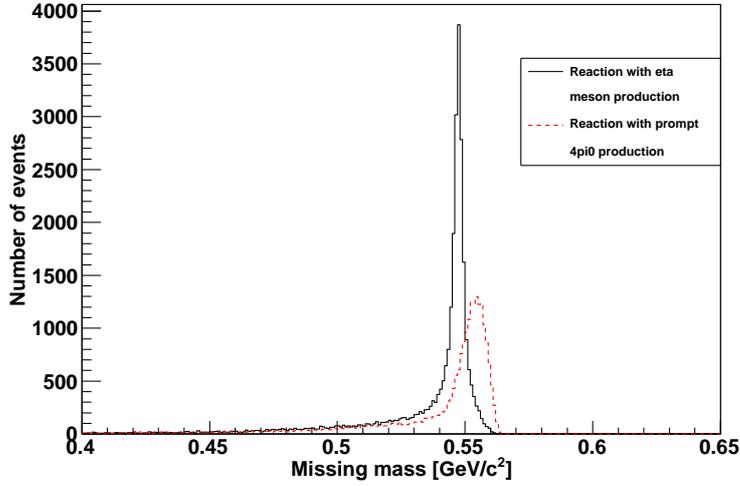}
	\caption{Missing mass for the $pp \to ppX$ reaction. Solid, black line denotes reaction where the $\eta$ meson is produced. Dashed, red line denotes prompt production of $4\pi^{0}$ in proton-proton collision. Both histograms are results of Monte Carlo simulations of $10^6$ events for $pp \to pp\eta \to pp4\pi^{0} \to pp8\gamma$ and $pp \to pp4\pi^{0} \to pp8\gamma$ reactions with $T_{beam} = 1.3\ GeV$. }
	\label{rys_missmass}
\end{figure}

$A_{\eta}$ can be established using Monte Carlo simulation as a ratio of observed events to produced events, $A_{\eta}=N_{\eta\ detected}/ N_{\eta\ produced}$. In the determination for a number of $N_{\eta\ detected}$ a missing mass technique is applied. The missing mass technique is essential to distinguish the decay of $\eta$ meson from prompt production of $4\pi^{0}$ in proton-proton collision (main background of $\eta \to 4\pi^{0}$). In Fig. \ref{rys_missmass} distribution of missing mass reconstructed from sample of simulated $pp \to pp\eta \to pp4\pi^{0}$ (solid, black line) and its main background $pp \to pp4\pi^{0}$ (dashed, red line) is plotted. These events obey all restrictions mentioned above in section \ref{sec:kinematics}. Full width at half maximum (FWHM) for $\eta$ signal is approximately $3\ MeV$. Taking as $N_{\eta \to 4\pi^{0}}$ a number of entries in the range of $6\ MeV$ (\mbox{$\pm$FWHM}) around the mean value one obtains that $A_{\eta}$ is equal to about $1.5\%$.

In order to determine uncertainty of number of observed events, information about background reactions is needed. The value of $N_{\eta \to 4\pi^{0}}$ can be in principle calculated as: 

\begin{equation}
 N_{\eta \to 4\pi^{0}} = N_{observed}-N_{background} = (N_{\eta \to 4\pi^{0}} + N_{background}) - N_{background}.
\label{tom_eq6}
\end{equation}

\noindent
Under assumption that events observed in experiment can be described by Poisson distribution and that
$N_{observed}$ as well as $N_{background}$ can be measured independently and that $N_{\eta \to 4\pi^{0}}$ is negligible compared to $N_{background}$, uncertainty of $N_{\eta \to 4\pi^{0}}$ equals to: 

\begin{equation}
 \sigma(N_{\eta \to 4\pi^{0}}) = \sqrt{2\cdot N_{background}}
 \label{tom_eq7}
\end{equation}

\noindent
Using equation \ref{tom_eq2} to describe $N_{background}$ we find:

\begin{equation}
	\sigma(N_{\eta \to 4\pi^{0}})= \sqrt{2 \cdot A_{4\pi^{0}} \cdot \sigma_{4\pi^{0}} \cdot L \cdot \Delta t},
	\label{tom_eq8}
\end{equation}

\noindent
where $L$ and $\Delta t$ denote the same variables as in equation \ref{tom_eq2}, $A_{4\pi^{0}}$ stands for acceptance of the detector for prompt four pions production and $\sigma_{4\pi^{0}}$ corresponds to cross section for $pp \to pp4\pi^{0}$ reaction. Employing equation \ref{tom_eq7} in equation \ref{tom_eq5} one obtains:

\begin{equation}
	\sigma(BR) = \sqrt{ \frac{2 \cdot A_{4\pi^{0}} \cdot \sigma_{4\pi^{0}} }{A_{\eta}^{2} \cdot \sigma_{\eta}^{2} \cdot L \cdot \Delta t}  }
	\label{tom_eq9}
\end{equation}

Using the same procedure for $A_{4\pi^{0}}$ as it is described above for $A_{\eta}$, one can obtain  $A_{4\pi^{0}} = 0.5\% $

Another variable in equation \ref{tom_eq2} which has not been mentioned yet is cross section. The cross section of the $\eta$ production ($\sigma_{\eta}$) is taken from the parameterization of experimental data \cite{Chiavassa, Pauly_art, Smyrski, Hibou, Flaminio:1984gr}. 

\begin{figure}[h]
	\centering
	\includegraphics[width=0.70\textwidth]{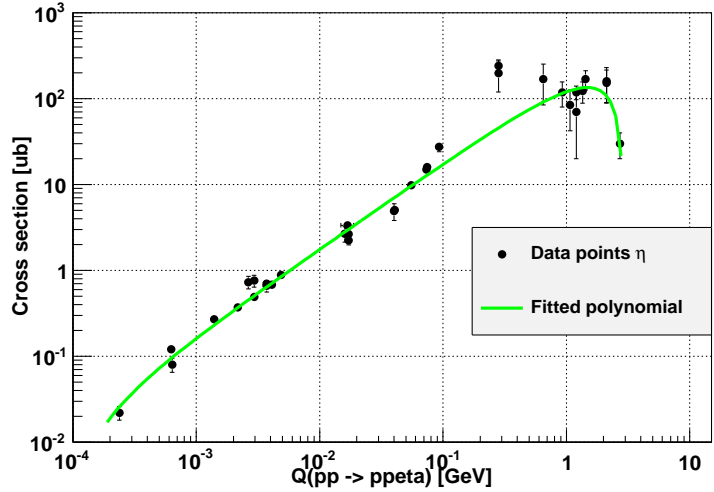}
	\caption{Cross sections for the $\eta$ meson production in proton-proton collision. Data are from Refs. \cite{Chiavassa, Pauly_art, Smyrski, Hibou, Flaminio:1984gr} and the superimposed line indicates result of the fitted polynomial $a+bQ+cQ^2+dQ^3$. }
	\label{rys_etaCross}
\end{figure}

To describe experimental  cross section for the $\eta$ meson production in proton-proton collision, third order polynomial ($a+bQ+cQ^2+dQ^3$) is fitted to data. The result of fitting is shown in Fig. \ref{rys_etaCross}. The values of polynomial coefficients are presented in Tab.~\ref{tab_polynomialCoeff}.  

\begin{table}[h]
	\centering
	\begin{tabular}{c|c}
		Coefficient & Value  \\
		\hline
		$a$ & $-0.0164 \pm 0.0040$  \\
		$b$ & $177.3 \pm 4.4$  \\
	  $c$ & $-55 \pm 17$  \\
	  $d$ & $-2.3 \pm 6.2$  \\
	\end{tabular}\\
		
	\caption{Coefficients of polynomial ($a+bQ+cQ^2+dQ^3$) fitted to cross sections for the $\eta$ meson production in proton-proton collision as shown in Fig. \ref{rys_etaCross}. Q is expressed in units of GeV. } 
	\label{tab_polynomialCoeff}
\end{table}

The cross section for the prompt pions production is taken from parameterization \ref{rppte_eq1} (see section \ref{reactions}) and its shape is shown as a green line in Fig.~\ref{rys_przekrojPiony}. 

Time ($\Delta t$) is a parameter.

\begin{figure}[h]
	\centering
	\includegraphics[width=0.70\textwidth]{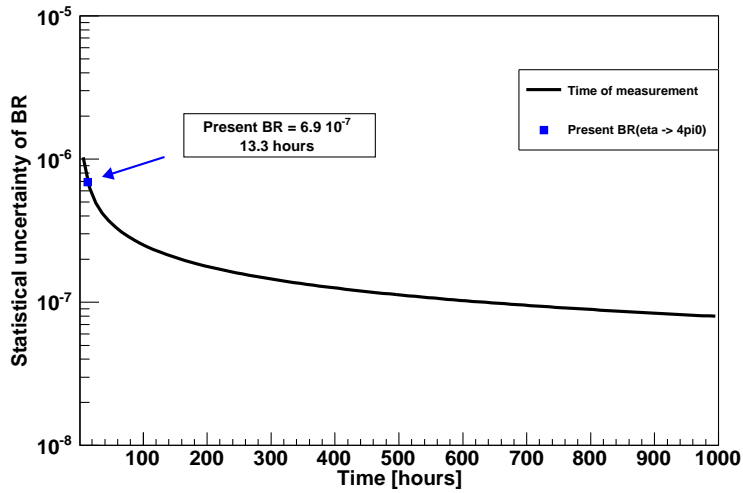}
	\caption{Statistical error of branching ratio as a function of time. Blue square presents value of upper limit of branching ratio \cite{CrystalBall}. Black line shows value of statistical error which can be achieved using WASA-at-COSY detector for measurement at $T_{beam}=1.3\ GeV$ with $L=10^{32}\ cm^{-2}s^{-1}$ and time $\Delta t$.}
	\label{rys_statError}
\end{figure}

Using equation \ref{tom_eq9}, statistical error of branching ratio as a function of time can be plotted and it is shown in Fig. \ref{rys_statError} under assumption that luminosity is constant and equal to $L=10^{32}\ cm^{-2}s^{-1}$. The result is obtained for proton-proton collision with beam kinetic energy of $T_{beam}=1.3\ GeV$. Upper limit of a branching ratio for the present value of the decay $\eta \to 4\pi^{0}$ (\ref{tom_eq1}) is marked and it can be achieved after 13.3 hours of measurement using \mbox{WASA-at-COSY} detector. However, this result depends strongly on the beam energy. Dependence on the excess energy is hidden in cross sections: $\sigma_{4\pi^{0}} = \sigma_{4\pi^{0}}(Q)$, $\sigma_{\eta} = \sigma_{\eta}(Q)$ and acceptances: $A_{\eta} = A_{\eta}(Q)$, $A_{4\pi^{0}} = A_{4\pi^{0}}(Q)$. Thus, by rewriting equation \ref{tom_eq9} for time of measurement as a function of $Q$, we find

\begin{equation}
	\Delta t = \frac{2 \cdot A_{4\pi^{0}}(Q) \cdot \sigma_{4\pi^{0}}(Q)}{A_{\eta}^{2}(Q) \cdot L \cdot \sigma^{2}(BR) \cdot \sigma^{2}_{\eta}(Q)}.
	\label{tom_eq10}
\end{equation}

\begin{figure}[h]
	\centering
	\includegraphics[width=0.70\textwidth]{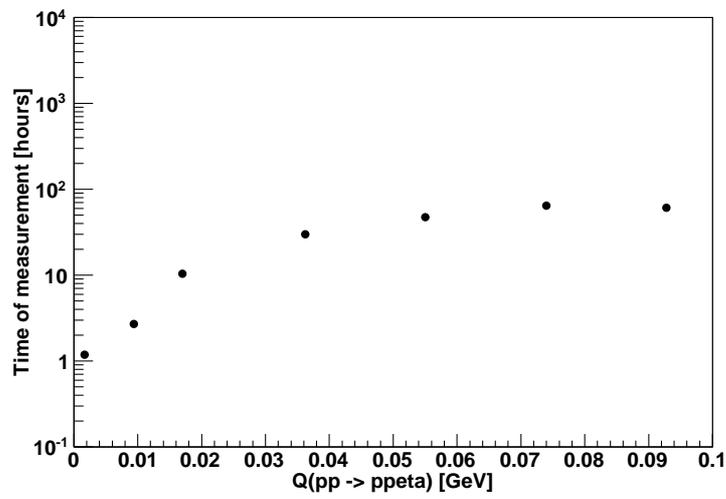}
	\caption{Relation between time of measurement needed to achieve statistical uncertainty of $6.9\cdot 10^{-7}$ and excess energy for the $\eta$ meson production in proton-proton collision assuming $L=10^{32}\ cm^{-2}s^{-1}$. }
	\label{rys_TimeVsQ}
\end{figure}

In order to find the optimum energy for the studies of the $pp \to pp\eta \to pp4\pi^{0}$ reaction, a time of measurement is plotted as a function of excess energy (Fig. \ref{rys_TimeVsQ}) with fixed value of $\sigma(BR)=6.9\cdot 10^{-7}$.

Fig.~\ref{rys_TimeVsQ} indicates that time of measurement needed to achieve a given accuracy of $BR(\eta \to 4\pi^{0})$  decreases with smaller values of excess energy for the $\eta$ meson production in proton-proton collisions.

%%%%%%%%%%%%%%%%%%%%%%%%%%%%%%%%%%%%%%%%%%%%%%%%%%%%%%%%%%%%%%%%%%%%%%%%%%%%%%%%%%%%%%%%%%%%
\newpage
\begin{center}\end{center}
\newpage
%%%%%%%%%%%%%%%%%%%%%%%%%%%%%%%%%%%%%%%%%%%%%%%%%%%%%%%%%%%%%%%%%%%%%%%%%%%%%%%%%%%%%%%%%%%%
\newpage
\thispagestyle{plain}

\fancyhf{} 
\fancyhead[LE,RO]{\textbf{\thepage}}
\fancyhead[RE]{\small\textbf{{Summary and conclusions}}}

\newpage

\section{Summary and conclusions}

\vspace{0.5cm}

The main aim of this thesis was to estimate the time of measurement of \\ $\eta \to 4\pi^{0}$ decay using \mbox{WASA-at-COSY} detector for which the present branching ratio (BR) limit can be improved. To this end a Monte Carlo simulations of\\ $pp \to pp\eta \to pp4\pi^{0} \to pp8\gamma$ reaction kinematics and simulations of \mbox{WASA-at-COSY} detector response were performed. 

The first step of signals analysis was the estimation of how many of $\eta$ decays could be reconstructed. For this purpose geometry and reconstruction efficiency of \mbox{WASA-at-COSY} detector for registration of gamma quanta from $\eta \to 4\pi^{0} \to 8\gamma$ decay were examined. The obtained result shows that only about 3\% of $\eta \to 4\pi^{0} \to 8\gamma$ events can be reconstructed. Resolution for reconstruction of eight gamma quanta invariant mass was established to $31\ MeV/c^{2}$ (standard deviation).

In the next step of analysis the detector acceptance for the $pp \to pp\eta \to pp4\pi^{0} \to pp8\gamma$ reaction ($A_{\eta}=N_{\eta\ detected}/ N_{\eta\ produced}$) was estimated to $A_{\eta} = 1.5\%$. A number of $N_{\eta\ detected}$ events was extracted from the missing mass distribution of $pp \to ppX$ reaction.

The last step of analysis was to evaluate time of measurement to obtain statistical uncertainty equal to the current upper limit of BR. For proton beam with kinetic energy of $T_{beam}=1.3\ GeV$ and luminosity of $L=10^{32}\ cm^{-2}s^{-1}$ the obtained time is equal to about $13$ hours. In addition, based on the parameterization of cross section for $pp \to pp\eta$ and $pp \to pp4\pi^{0}$ reactions the time needed for measurement was also studied as a function of excess energy for $\eta$ meson production in proton-proton collision. From the determined dependence it was concluded that the best conditions for the registration of the $pp \to pp\eta \to pp4\pi^{0} \to pp8\gamma$ reaction with the \mbox{WASA-at-COSY} detector are for beam energies close to the kinematical threshold for $pp \to pp\eta$ reaction. 

Obtained results indicate that statistical uncertainty at level of present branching ratio limit is  achievable with the \mbox{WASA-at-COSY} experiment. However, its significant improvement will require long term measurement. In order to improve the current limit by about one order of magnitude about 41 days of the measurement at $T_{beam}=1.3\ GeV$ and luminosity of $L=10^{32}\ cm^{-2}s^{-1}$ is needed corresponding to the production of about $10^{9}$ $\eta$ mesons.

Up to now almost $4\cdot 10^{8}$ decays of the $\eta$ meson have been collected using \mbox{WASA-at-COSY} detector.

Consequently, the outcome of this thesis demonstrates that it is worth to search in these data for the $\eta \to 4\pi^{0}$ decay. 

%%%%%%%%%%%%%%%%%%%%%%%%%%%%%%%% APPENDIX %%%%%%%%%%%%%%%%%%%%%%%%%%%%%%%%%%%%%%%%%%%%%%%%%%%%%%%%%%%%%%%%%
\newpage
\begin{center}\end{center}
\newpage
%%%%%%%%%%%%%%%%%%%%%%%%%%%%%%%%%%%%%%%%%%%%%%%%%%%%%%%%%%%%%%%%%%%%%%%%%%%%%%%%%%%%%%%%%%%%
\newpage
\thispagestyle{plain}

\appendix 
  \section{Kinematical Fit}
  
\vspace{0.5cm}  
  
\pagestyle{fancy}
\fancyhf{}
\fancyhead[LE,RO]{\textbf{\thepage}}
\fancyhead[RE]{\small\textbf{{Appendix\ \ Kinematical Fit  }}}

\begin{figure}[h!]
	\centering
	\includegraphics[width=0.7\textwidth]{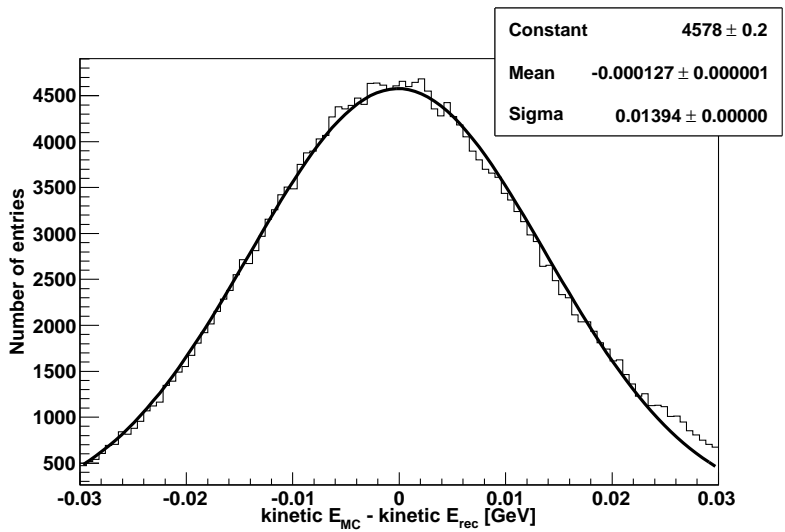}
	\caption{Difference between true and  reconstructed kinetic energy of gamma quanta determined from signals in the electromagnetic calorimeter. In order to determine standard deviation, Gaussian function is fitted. }
	\label{rys_PhEnergy}
\end{figure}

Distribution of $\eta$ meson mass reconstructed as invariant mass of eight gamma quanta is quite broad (see  Fig.~\ref{rys_invMeta}). In order to improve this resolution, kinematical fit technique is applied. Kinematical fit varies measured quantities for particles within the measurement uncertainty to match constraints of working hypothesis and optimize precision of the studied variables. In this case, as constraints it is imposed that invariant mass of every pair of gamma quanta matched by routine (see subsection \ref{subsec:pions_reco}) should be equal to neutral pion mass. Quality of fit is measured by $\chi^{2}$ calculation. Varied quantities for protons and gamma quanta are kinetic energy $E_{k}$, polar angle $\theta$ and azimuthal angle $\phi$. Uncertainty of these quantities are obtained on a basis of the Monte Carlo simulations by the comparison of distribution of the difference between true and reconstructed values ($q_{MC} - q_{rec}$). As an example of this kind of distribution, kinetic energy of gamma quanta is shown in Fig. \ref{rys_PhEnergy}.

All standard deviations obtained from Gaussian function fitting for each variable are presented in Tab.~\ref{sigmaKFit_tab}.

\begin{table}[h!]
	\centering
	\begin{tabular}{c|c|c}
		\hline 
		Particles  & Parameter & $\sigma$  \\ 
		\hline
		Protons & $E_{kin}$ & $8.8\ MeV$ \\
		\cline{2-3}
	  & $\theta$ & $0.2\ deg$  \\
	  \cline{2-3}
	   & $\phi$ & $2.0\ deg$  \\
	  \hline
	  Photons & $E_{kin}$ & $13.9\ MeV$  \\
	  \cline{2-3}
	  & $\theta$ & $1.8\ deg $  \\
	  \cline{2-3}
	  & $\phi$ & $2.4\ deg$  \\
	\end{tabular}

	\caption{Standard deviations used in kinematical fit program.} 
	\label{sigmaKFit_tab}
\end{table}

\begin{figure}[h!]
	\centering
	\includegraphics[width=0.8\textwidth]{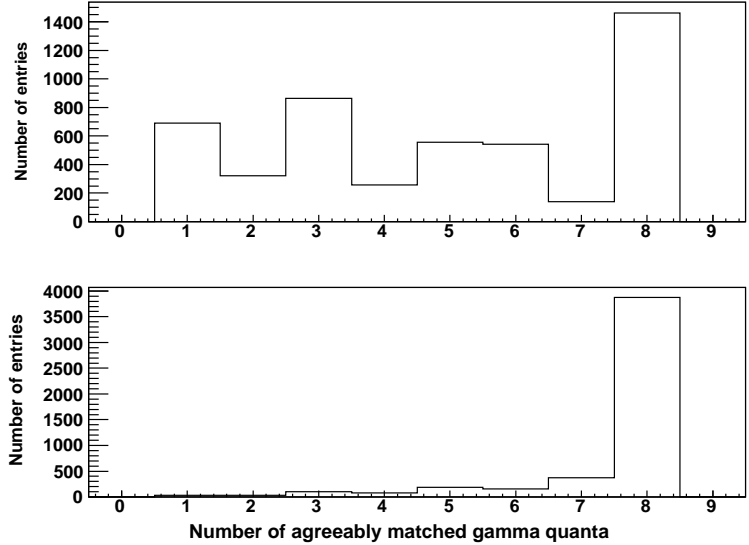}
	\caption{Number of gamma quanta correctly matched into pairs by the used algorithm. The upper histogram shows gamma quanta for the best set chosen by the algorithm. The bottom one presents the choice with the biggest number of properly assigned gamma quanta from four best possibilities found by the matching algorithm (see subsection \ref{subsec:pions_reco}). }
	\label{rys_PhotonsOK}
\end{figure}

Kinematical fit for input data mentioned above did not work well. Therefore, efficiency of gamma quanta matching algorithm is performed. For the best set of gamma quanta according to the routine, a distribution of number of properly assigned gamma quanta is presented in upper histogram in Fig. \ref{rys_PhotonsOK}. As one can notice only in about $1/3$ of reactions gamma quanta are fully properly assigned.

For comparison purposes, beside the best set of gamma quanta according to the algorithm, three additional sets are checked.  Additional sets are these with a larger sum of differences between pion mass and masses of matched pairs (see subsection~\ref{subsec:pions_reco}). The chosen set taken into account in Fig.~\ref{rys_PhotonsOK} is the one for which number of correctly matched gamma quanta is the biggest. The result of this test is shown in bottom histogram in Fig. \ref{rys_PhotonsOK}. One can observe a significant improvement in gamma quanta assignment. 

Signals reconstructed from Monte Carlo simulation can be compared with true values, however in experiment other criterion of properly matched gamma quanta is needed. In experiment $\chi^{2}$ value from the kinematical fit can be used. 

\begin{figure}[h!]
	\centering
	\includegraphics[width=0.8\textwidth]{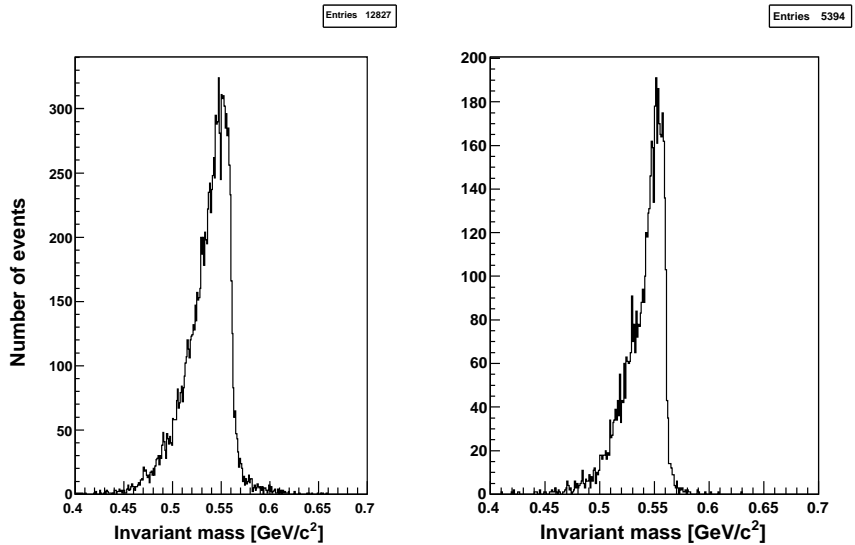}
	\caption{The right hand side histogram shows invariant mass of gamma quanta for the best set of the gamma quanta found by the matching algorithm. Histogram on the left hand side presents the choice with the $\chi^{2}$ value closest to the expected value of $\chi^{2}$ distribution of $pp \to pp\eta \to pp4\pi^{0} \to pp8\gamma$ reaction.}
	\label{rys_invMetaKF}
\end{figure}

Similar comparison to this above is done for events with $\chi^{2}$ value as criterion of proper gamma quanta assignment. In Fig.~\ref{rys_invMetaKF}, right hand side histogram shows invariant mass of eight gamma quanta after kinematical fit taking into account only the best outcome of the matching algorithm. Histogram in the left panel shows result after kinematical fit taking the outcome with the best $\chi^{2}$ out of four best possibilities found by the matching algorithm. Set assumed to be the best is the one with $\chi^{2}$ value closest to the expected value of $\chi^{2}$ distribution of $pp \to pp\eta \to pp4\pi^{0} \to pp8\gamma$ reaction. 

FWHM of the distribution in Fig.~\ref{rys_invMetaKF} (left) is equal to $40\ MeV/c^{2}$ ($\sigma = 17\ MeV/c^{2}$) and it is almost two times smaller compared to one obtained in subsection~\ref{subsec:eta_reconstruction} ($\sigma = 31\ MeV/c^{2}$). One can see that number of entries is more than two times higher in case when four sets of gamma quanta are checked. However, still low statistics and non-gaussian distributions of both invariant mass histograms in Fig.~\ref{rys_invMetaKF} indicate that kinematical fit does not work satisfactorily for that input data. Therefore, further studies will be continued in the future. 

%%%%%%%%%%%%%%%%%%%%%%%%%%%%%%%%%%%%%%%%%%%%%%%%%%%%%%%%%%%%%%%%%%%%%%%%%%%%%%%%%%%
\newpage
\begin{center} \end{center}
\newpage
%%%%%%%%%%%%%%%%%%%%%%%%%%%%%%%%%%%%%%%%%%%%%%%%%%%%%%%%%%%%%%%%%%%%%%%%%%%%%%%%%%%

\newpage
\thispagestyle{empty} 
\pagestyle{fancy}
\fancyhf{}
\fancyhead[LE,RO]{\textbf{\thepage}}

\begin{center}
\Large \textbf{Acknowledgment}
\end{center}

\vspace{1.0cm}

\selectlanguage{polish}

At first I wish to thank Prof.~dr~hab.~Paweł~Moskal for his guidance during writing this diploma thesis. He has supported my self-development and showed me how interesting physics can be. \\

I would like to thank Prof.~dr~hab.~Bogusław Kamys for allowing me to prepare this thesis in the Nuclear Physics Department of the Jagiellonian University.\\

I am greatly indebted to Master of Science Małgorzata Hodana, Master of Science Wojciech Krzemień and Master of Science Marcin Zieliński for teaching me basics of WASA-MC, RootSorter, the time spent with me and all the help in solving problems which came up during the process. \\

I wish to thank Dr Aleksandra Wrońska for teaching me basics of ROOT and all the help in solving problems related to it. \\

I thank Doc.~dr~hab.~Andrzej Kupść for a scientific contribution.\\

I would like to thank PD.~dr~hab.~Frank Goldenbaum for reading the manuscript.\\

I am grateful to all the \mbox{WASA-at-COSY} members for their help and friendly atmosphere.\\

I would like to thank WIKING programme for the opportunity of doing practice in Institute of Nuclear Physics, Polish Academy of Sciences.\\

I also would like to thank my colleagues: Izabela Balwierz, Szymon Niedźwiecki, Master of Science Michał Silarski, Master of Science Magdalena Skurzok and Master of Science Tomasz Twaróg for a scientific contribution and a great time spent together. Especially Szymon for our get-togethers and discussions during the whole time of studies and Tomasz for reading part of manuscript.\\

I thank very much my friends Ada Umińska and Monika Josiekova for being close to me and enormous support during last years.\\

Last but not least, I want to thank my parents and my brother for their love and help during my entire life and I would like to thank my brother's wife, Agata, for language corrections in manuscript.

%%%%%%%%%%%%%%%%%%%%%%%%%%%%%%%%%%%%%%%%%%%%%%%%%%%%%%%%%%%%%%%%%%%%%%%%%%%%%%%%%%

\newpage
\pagestyle{fancy}
\fancyhf{} 
\fancyhead[LE,RO]{\textbf{\thepage}}

\newpage
\thispagestyle{plain}

\selectlanguage{english}

\end{document}